\documentclass[%
 reprint,
 amsmath,amssymb,
 aps,
]{revtex4-2}
\usepackage{comment}

\usepackage{float}
\usepackage{braket}
\usepackage{graphicx}
\usepackage{dcolumn}
\usepackage{bm}

\usepackage{empheq}

\begin{document}

\preprint{APS/123-quantum electrodynamics}

\title{A general framework for interactions \\ between electron beams and quantum optical systems}

\author{Jakob~M.~Grzesik}
\email{yakgrz@stanford.edu}
\author{Aviv~Karnieli}
\email{karnieli@technion.ac.il}
\author{Charles~Roques-Carmes}%
\email{chrc@stanford.edu}   
\author{Dylan~S.~Black}
\author{Trung~Kiên~Lê}
\author{Olav Solgaard}
\author{Shanhui Fan}
\author{Jelena Vu\v{c}kovi\'{c}}

\affiliation{%
E. L. Ginzton Laboratories, Stanford University, 348 Via Pueblo, Stanford, CA 94305
}%

\date{\today}

\begin{abstract}
We provide a theoretical framework to describe the dynamics of a free-electron beam interacting with quantized bound systems in arbitrary electromagnetic environments. This expands the quantum optics toolbox to incorporate free-electron beams for applications in highly tunable quantum control, imaging, and spectroscopy at the nanoscale. The framework recovers previously studied results and shows that electromagnetic environments can amplify the intrinsically weak coupling between a free-electron and a bound electron to reach previously inaccessible interaction regimes. We leverage this enhanced coupling for experimentally feasible protocols in coherent qubit control and towards the nondestructive readout and projective control of the electron beam’s quantum-number statistics. Our framework is broadly applicable to microwave-frequency qubits, optical nanophotonics, cavity quantum electrodynamics, and emerging platforms at the interface of electron microscopy and quantum information.
\end{abstract}

\maketitle


\section{\label{sec:Intro}Introduction}
Engineering coherent interactions between quantized modes of light and matter lies at the core of emerging quantum technologies. Advances in nanofabrication have enabled semiconductor devices with control over photonic modes and materials processing has introduced various qubit candidates, such as superconducting circuits, semiconductor spins and color center defects. The co-integration of spin qubits with photonics offers an experimental platform for quantum simulation~\cite{altman2021quantum, georgescu2014quantum, lukin2025mesoscopic}, quantum communication~\cite{hermans2022qubit, pompili2021realization, knaut2024entanglement} and sensing \cite{pelucchi2022potential, degen2017quantum, turunen2022quantum, hepp2019semiconductor, biswas2025quantum}. Concurrent advances in electron beam generation have enabled tunable coherent interactions between electron beams and photonic modes~\cite{carmes2023, doi:10.1021/acsphotonics.5c00585, henke2021integrated, kfir2020controlling}, allowing for beam acceleration~\cite{sapraACHIP,litos2014high, shiloh2022miniature}, field imaging~\cite{fishman2023imaging, zhu2025nanometre, polman2019electron, shibata2017direct}, and microscopy~\cite{barwick2009photon, park2010photon, garcia2010multiphoton}. This interaction has even been explored experimentally in the quantum regime, with observations of coherent interaction between quantized energy states of free electrons and photonic modes~\cite{kfir2021optical, feist2015quantum}. 

Initial theoretical work on the resonant interaction between a dipolar two level system (TLS) and a bunched electron beam, dubbed ``free-electron bound-electron resonant interaction" (FEBERI), has garnered excitement as a resource for nanoscale quantum control~\cite{gover2020free} and entanglement generation~\cite{zhao2021quantum} mediated by the beam's electric field and qubit's electric dipole moment. In addition to theoretical proposals for coherent interactions between a free-electron's magnetic field and spin qubits~\cite{Haslinger_2024, PhysRevResearch.3.023247, even2025spin}, recent experiments have demonstrated promising steps towards quantum control and spin sensing with electron beams~\cite{kolb2025coherentdrivingquantummodulated, grzesik2025quantumsensingelectronbeams}. However, the inherently low coupling strength between free electrons and bound electrons remains an outstanding limitation for even these advanced experimental systems, often requiring probing distances and electron beams that are difficult to realize. Motivated by these challenges and frontiers, we explore methods to enhance these interactions and propose experimentally feasible protocols to study the effects of such an interaction on both the electron beam and spin-qubit system.

Here, we introduce a theoretical framework that unifies interaction models between a bound qubit, an arbitrary electromagnetic environment, and an electron beam, which respectively act as three distinct quantum systems: a discrete TLS, a unidirectionally infinite ladder, and a bidirectionally infinite ``quantum rotor"\cite{ruimy2025free}. By applying this formalism to a free electron beam interacting with a bound spin qubit in a microwave cavity, we may enhance the inherently weak coupling strength between the free-electron beam and bound-electron qubit towards useful regimes in quantum control and sensing. Such an enhancement enables the precise readout and non-destructive control of the electron beam's quantum number statistics. While this work focuses on the microwave regime, we emphasize that our formalism readily applies across the electromagnetic spectrum.
\vspace{-0.03cm}
\section{Results}
We consider the following system: an electron beam, traveling at velocity $v$ (determined by its energy) flies by a qubit embedded in an environment with a dyadic Green's function $\mathbf{G}(\mathbf{r},\mathbf{r'},\omega)$. The free-electron--qubit interaction is mediated by the electromagnetic field's photons and is described by an effective Hamiltonian within a Markovian master equation approach, where the photonic degrees of freedom are adiabatically eliminated, as detailed in the SI Section I. Remarkably, for free space and a cavity-mediated interaction in the dispersive regime, the effective Hamiltonian becomes approximately hermitian, so we can focus on unitary dynamics. 

\begin{figure*}
\includegraphics[width=0.9\textwidth]{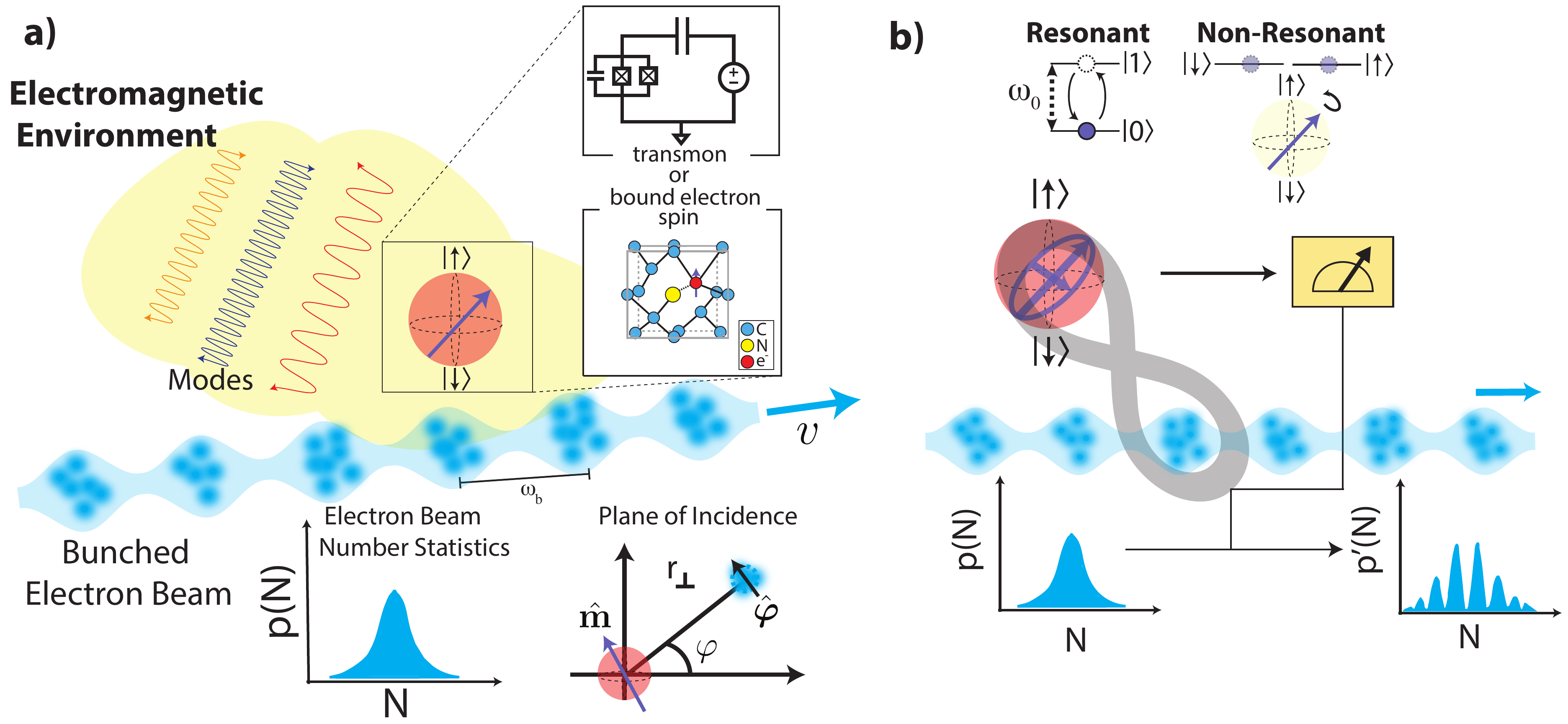}
\caption{\textbf{General electron-beam spin qubit interaction }\textbf{a.} System Schematic: an electron beam with an electron number distribution $p(n)$ and engineered wavefunction interacts with a qubit in an arbitrary electromagnetic environment. \textbf{b.} Effects of the interaction on both the spin qubit (for resonant and non-resonant cases, top) and the electron beam (bottom). The interaction entangles qubit precession and the electron beam's quantized number distribution. Projective readout of the qubit thus modifies the electron beam distribution (bottom right).}
\label{fig:fig1}
\end{figure*}

\textit{Free-space interaction.} We define electron beam propagation along the $z$ axis, and position the spin qubit at the origin, depicted in Figure~\ref{fig:fig1}(a). The plane of incidence sets impact parameter $r_\perp$ and angle $\varphi$, denoted by a 2D vector $\mathbf{r}_\perp$. The spin's magnetic moment, aligned along $\hat{\mathbf{m}}$, has a resonance frequency $\omega_0$, and spin lowering (raising) operator $\hat{\sigma}$ ($\hat{\sigma}^{\dagger}$). The structure of the free space Green's function in the near-field~\cite{novotny2012principles} simplifies the interaction Hamiltonian. Applying the rotating wave approximation yields an effective Hamiltonian
\begin{equation}
    \hat{H}_{\text{int}}(t) = -\frac{g\mu_B}{\hbar} \left[\hat{\mathbf{S}}^{(+)}(t)\cdot\hat{\mathbf{B}}_{\mathrm{el}}^{(-)}(t) +\hat{\mathbf{S}}^{(-)}(t)\cdot\hat{\mathbf{B}}_{\mathrm{el}}^{(+)}(t)  \right] 
    \label{eq: general_int}
\end{equation}
where $\hat{\mathbf{S}}^{(-)}(t)=e^{-i\omega_0t}(\hbar /2)\hat{\mathbf{m}} \hat{\sigma}$ is the spin operator, $\mu_B$ is the Bohr magneton, $g\simeq 2$ is the $g$-factor, and  $\hat{\mathbf{B}}_{\mathrm{el}}^{(+)}(t)=\hat{\mathbf{B}}_{\mathrm{el}}^{(+)}(\mathbf{0},t)$ is the positive-frequency magnetic field induced by the electron at the spin's position
\begin{equation}
    \hat{\mathbf{B}}_{\mathrm{el}}^{(+)}(t)=-\hat{\boldsymbol{\varphi}
} \frac{\mu _0 e}{2\pi r_{\perp}} \int_0^\infty d\omega e^{-i\omega t} \left(\frac{\omega}{v}r_{\perp }\right) K_1\left(\frac{\omega}{v} r_{\perp }\right)\hat{b}_{\frac{\omega}{v}}
\label{eq:field_td}
\end{equation}
Here, $\hat{\boldsymbol{\varphi}}$ is a unit vector tangential to a circle of radius $r_{\perp}$ around the origin and at the position of the electron beam (see Fig. 1a), $K_1(x)$ is the first-order modified Bessel function of the second kind, and $\hat{b}_{q}$ the electron ladder operator~\cite{ruimy2025free, kfir2019entanglements, feist2015quantum}, which lowers the electron momentum by a recoil $q$. The spin position and orientation relative to the electron beam directly impact $\hat{\mathbf{B}}_{\mathrm{el}}^{(+)}$ through $\hat{\boldsymbol{\varphi}}$.

We obtain the scattering matrix $\hat{S}=\mathcal{T}\exp\left[-i/\hbar\int_{-\infty}^{\infty}\hat{H}_{\text{int}}(t)dt\right]$ using the first-order Magnus expansion~\cite{magnus1954exponential}
\begin{equation}
    \hat{S} = \exp\left[-i \phi_0\left(\hat{\mathbf{m}}\cdot\hat{\boldsymbol{\varphi}
}\right)\hat{\sigma}\hat{b}^{\dagger}_{\frac{\omega_0}{v}} -i \phi_0\left(\hat{\mathbf{m}}^*\cdot\hat{\boldsymbol{\varphi}
}\right)\hat{\sigma}^{\dagger}\hat{b}_{\frac{\omega_0}{v}} \right],
\label{eq:scatter_gen}
\end{equation}
where
\begin{equation}
      \phi_0 = \frac{\mu_B\mu_0e}{2\pi r_{\perp}\hbar} \left(\frac{\omega_0}{v}r_{\perp }\right) K_1\left(\frac{\omega_0}{v} r_{\perp }\right)\simeq \frac{\mu_B\mu_0e}{2\pi r_{\perp}\hbar} 
    \label{eq:impact_param}
\end{equation}
is the dimensionless free-space coupling constant between the electron beam and the spin~\cite{grzesik2025quantumsensingelectronbeams}, using approximations valid for deep-subwavelength impact parameters $r_{\perp}\ll \lambda$. Within the formalism of Eq. (1), the spin undergoes incoherent decay into free space at a rate $\Gamma_{\mathrm{sp}}$, so we shall assume that the scattering interaction happens on timescales much shorter than this lifetime.

Because the free space interaction strength $\phi_0$ is inversely proportional to $r_{\perp}$, it typically varies between $10^{-6}$ to $10^{-10}$ for impact parameters between $1~\mathrm{nm}$ to tens of $\mu\mathrm{m}$. Thus, macroscopically large currents are required to measurably affect both single qubits and spin ensembles. Experimentally, electron-beam-induced damage to the sample restricts impact parameters to larger values ($\sim10~\mu\mathrm{m}$)~\cite{grzesik2025quantumsensingelectronbeams}, limiting the coupling strength. 

\textit{Cavity-mediated interaction.} However, we may overcome such limitations by placing the qubit inside a microwave cavity, wherein the cavity field mediates a coherent interaction between the qubit and electron beam. For strongly coupled semiconductor spins or superconducting qubits, this interaction can reach a regime where the qubit can sense the beam's number statistics and entangle with the total electron number. We now consider a microwave cavity with linewidth $\gamma$, and resonant modes with wavenumber and frequency $q_j,\omega_j$. The electric and magnetic vacuum fields of the cavity modes are $\boldsymbol{\mathcal{E}}_j(\mathbf{r})$ and $\boldsymbol{\mathcal{B}}_j(\mathbf{r})$, respectively.

In order to avoid enhancing spontaneous energy decay from both the electron beam and spin excitations into the photonic bath, we consider the dispersive regime, commonly employed in cavity quantum electrodynamics~\cite{PhysRevA.74.042318, schuster2007resolving}. The interacting particles are far off-resonant from the cavity modes, such that the detuning $\Delta = \omega_m-\omega_0$ from the closest resonance (at a particular $j=m$) satisfies $\Delta\gg\gamma$. For the free electron, the frequency scale is determined by its phase-matching condition, manifesting a phase mismatch given by $q_m v = \omega_0 = \omega_m-\Delta$. However, since the free-electron's phase-matching bandwidth, inversely proportional to its time of flight through the cavity, sets the effective linewidth~\cite{dahan2020resonant}, we further assume a slow-enough electron, such that its interaction time with the cavity $T=L/v$ (where $L$ is the cavity length) satisfies $\Delta T\gg 2\pi$. Electron beams with keV to sub-keV energies meet these conditions~\cite{karnieli2023jaynes, karnieli2023quantum}, but are still swift enough to satisfy $\gamma T\ll 1$, such that the cavity lifetime is still much longer than the interaction time. This gives the effective interaction Hamiltonian (see SI section IB)
\begin{equation}
    \hat{H}_{\text{int}} = \hbar \frac{gg_{\mathrm{el}}^*}{\Delta} \hat{\sigma}\hat{b}^{\dagger}_{\frac{\omega_0}{v}} +\hbar \frac{g^*g_{\mathrm{el}}}{\Delta}\hat{\sigma}^{\dagger}\hat{b}_{\frac{\omega_0}{v}},
    \label{eq:H_int_cav}
\end{equation}
where $g\in\{g_{\mathrm{sp}},  g_{\mathrm{d}}\}$, with $g_{\mathrm{sp}}=\mu_B \hat{\mathbf{m}}\cdot\boldsymbol{\mathcal{B}}_m(\mathbf{r_{\mathrm{qu}}})/\hbar$, $g_{\mathrm{d}}=\mathbf{d}\cdot\boldsymbol{\mathcal{E}}_m(\mathbf{r}_{\mathrm{qu}})/\hbar$, and $g_{\mathrm{el}} = ev\hat{\mathbf{z}}\cdot\boldsymbol{\mathcal{E}}_m(\mathbf{r}_{\mathrm{el}})/\hbar\omega_m$, are the spin-magnetic field ($g_{\mathrm{sp}}$), dipole-electric field ($g_{\mathrm{d}}$), and free-electron-electric field ($g_{\mathrm{el}}$) coupling rates, respectively. The cavity-mediated coupling in $\hat{H}_{\mathrm{int}}$ depends directly on the spatial mode structure of the cavity field at the qubit and free-electron positions $\mathbf{r}_{\mathrm{qu}},\mathbf{r}_{\mathrm{el}}$, rather than decreasing with their relative distance. The corresponding scattering matrix is
\begin{equation}
    \hat{S} = \exp\left[-i \frac{gg_{\mathrm{Q}}^*}{\Delta}\hat{\sigma}\hat{b}^{\dagger}_{\frac{\omega_0}{v}} -i \frac{g^*g_{\mathrm{Q}}}{\Delta}\hat{\sigma}^{\dagger}\hat{b}_{\frac{\omega_0}{v}} \right]
    \label{eq:scatter_cav}
\end{equation}
where $g_{\mathrm{Q}}=g_{\mathrm{el}}T$ is the dimensionless coupling constant between the free electron and the cavity mode~\cite{polman2019electron, di2019probing}. As in the free space interaction, we need to also consider the incoherent decay of the qubit and the electron into the cavity. In the strongly detuned regime $\Delta \gg \gamma$ these decay rates are strongly suppressed, given approximately by $\Gamma_{\mathrm{qu}}\simeq (\gamma/2)|g/\Delta|^2$ and $\Gamma_{\mathrm{el}}\simeq 4|g_{\mathrm{el}}|^2/\Delta ^2 T$. To safely neglect them, we require that the qubit decays at a much slower rate than cavity photons, $\Gamma_{\mathrm{qu}}\ll\gamma$, and that the electron decay rate satisfies $\Gamma_{\mathrm{el}} T\ll 1$. These requirements boil down to $g,g_{\mathrm{el}}\ll\Delta$. Therefore, the characteristic strength for a cavity-mediated interaction between a free electron and spin is
\begin{equation}
    \phi_{\mathrm{cav}} = |g_{\mathrm{Q}}| \frac{|g|}{\Delta}.
    \label{eq:impact_cav}
\end{equation}
The cavity-mediated interaction is enhanced through the electron-cavity coupling $g_{\mathrm{Q}}$, which may be orders of magnitude larger than the free space electron-qubit coupling. Indeed, recent advances in free-electron--light interactions have shown that the strong coupling regime where $|g_{\mathrm{Q}}|=1$ is attainable in the optical domain~\cite{ZhexinBound}, and, theoretically, is also realizable and surpassable in the microwave regime~\cite{xie2025maximal}. 

Whether $\phi_{\mathrm{cav}}$ is enhanced (relative to the free-space case) depends directly on the qubit's coupling strength to the cavity modes, which may differ dramatically. For example, NV center ensembles in microwave cavities typically couple at $\sim$MHz~\cite{eisenach2021cavity} rates, while for atoms it is in the 10's to 100's of MHz~\cite{samkharadze2018strong, mi2018coherent}, and superconducting qubits can readily reach rates greater than 100's of MHz~\cite{krasnok2024superconducting, majer2007coupling, upadhyay2021robust}. For strongly coupled qubits, such as a superconducting qubit or a semiconductor spin, we may consider a qubit with $|g|=0.1\Delta$ and $|g_Q|=1$. This setting leads to an interaction strength of $\phi_{\mathrm{cav}}=0.1$, multiple orders of magnitude stronger than the free-space parameter $\phi_0$ evaluated earlier, enabling a regime that accounts for the quantum nature of the electron beam.

\textit{Entanglement with electron number.} The scattering matrices Eqs.~(\ref{eq:scatter_gen})and~(\ref{eq:scatter_cav}) describe a coherent resonant interaction between a free electron and a microwave qubit, with resonant frequency $\omega_0$. This scattering matrix recovers the FEBERI regime~\cite{gover2020free}, wherein a longitudinal modulation of the electron beam strengthens the interaction, observable by writing $\hat{b}_{\omega_0/v} = \int dz e^{-i\omega_0z/v}\hat{n}(z)$. Here, $\hat{n}(z)$ denotes the electron number density operator. Conventionally, FEBERI models consider high (e.g. optical) frequencies $\omega_0$, where $\hat{b}_{\omega_0/v}$ imparts a measurable recoil $\omega_0/v$ on the electron, correlated with an emitter excitation or decay. A spectrally-narrow and unmodulated electron beam can then demonstrate entanglement between the electron energy and emitter excitation~\cite{gover2020free, zhao2021quantum, ruimy2021toward}.


An interesting observation emerges for electron bunches pulsed at low (e.g., microwave) frequencies. For a bunch duration much smaller than $1/\omega_0$, the electron pulse is spectrally broader than the imparted recoil, making the electronic state before and after the interaction approximately identical. In this readily accessible regime, we take $\omega_0\to 0$ and consider the operator $\hat{b}_0=\hat{b}_0^{\dagger}=\int dz \hat{n}(z) \equiv \hat{N}$, where $\hat{N}$ denotes the bunch's total electron number operator. While this picture becomes exact for degenerate spin manifolds where $\omega_0\equiv 0$, it remains an excellent approximation for $\mathrm{GHz}$ resonance frequencies and sub-picosecond electron pulses available in ultrafast electron microscopes~\cite{di2019probing, ischenko2014ultrafast, plemmons2015probing, polman2019electron}. This produces the scattering matrix
\begin{equation}
    \hat{S} = \exp\left(-i \phi \left(e^{i\alpha} \hat{\sigma} + e^{-i\alpha} \hat{\sigma}^{\dagger}\right) \hat{N} \right),
    \label{eq: NonRes}
\end{equation}
where $\phi\in\{\phi_0, \phi_{\mathrm{cav}}\}$ is the coupling strength and $\alpha$ denotes the argument of the complex-valued couplings in Eqs.~(\ref{eq:scatter_gen}) and~(\ref{eq:scatter_cav}), dependent on electron beam transverse position (here onwards, we set $\alpha=0$). This operator produces a unique quantum interaction where the spin's precession becomes \textit{entangled} with the beam's electron number. 

\begin{figure}
\includegraphics[width=0.45\textwidth]{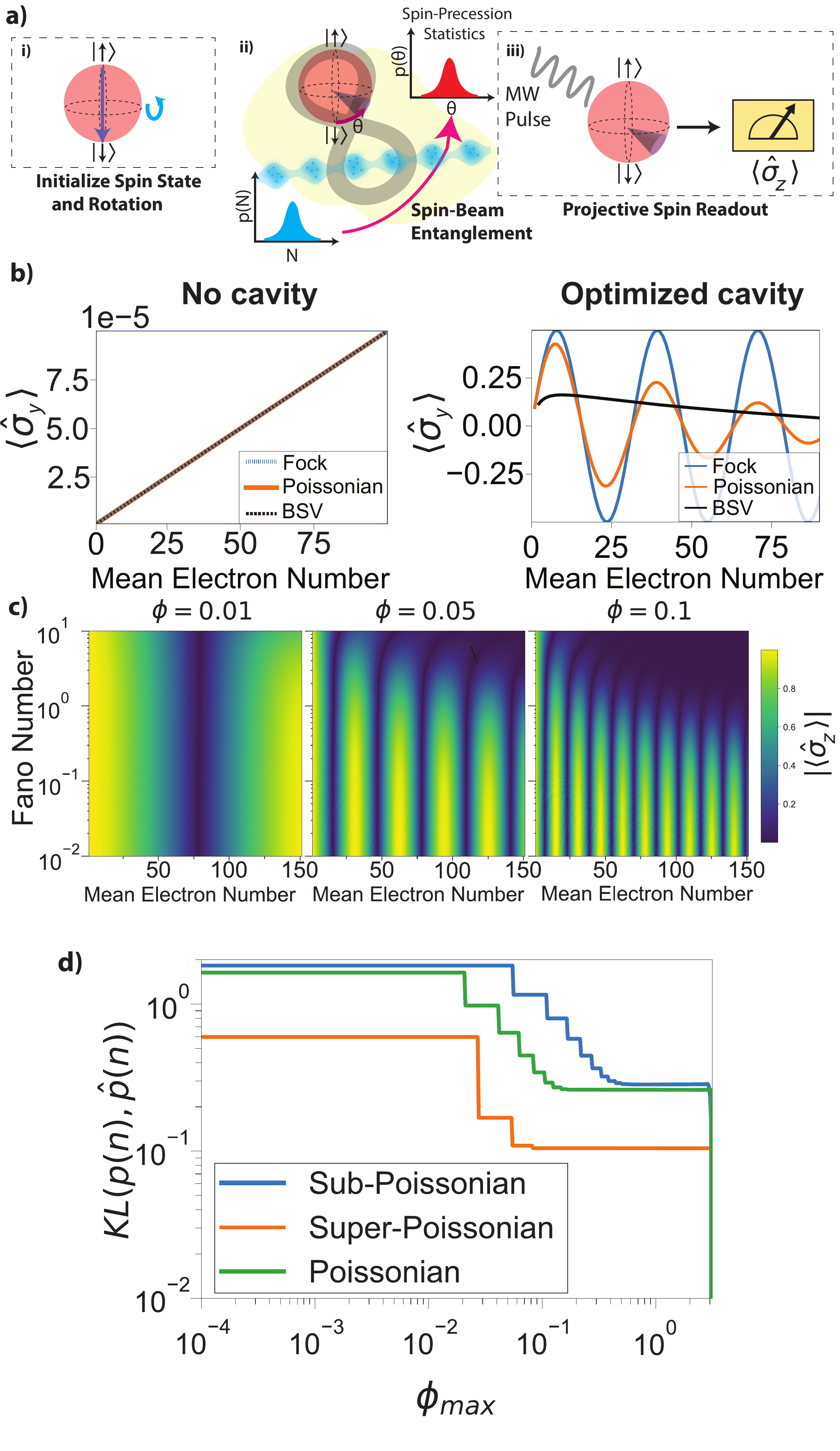}
\caption{\textbf{Electron Statistics Discrimination and Determination }\textbf{a.} Three stage experiment: (i): Spin initialization in $\ket{\downarrow}$ state; (ii): Electron beam-spin interaction, entangling spin evolution with electron beam number distribution; (iii): Spin readout of $\hat{\sigma}_z$, with an optional rotation prior to readout to select a measurement axis. \textbf{b. }$\langle \hat{\sigma}_y\rangle$ readout for different effective interaction strengths $\phi_0$, for free space (left) and an optimized cavity with localized fields (right). \textbf{c. }$\langle \hat{\sigma}_z\rangle$ for various distributions, characterized by the mean electron number $\mu$, and the Fano number $F \equiv \sigma_N^2/\mu$ with different interaction strengths. \textbf{d. }Kullback-Leibler (KL) divergence between the true electron number distribution $p(n)$ and the recovered distribution $\hat{p}(n)$ using the inverse Fourier Transform, as a function of maximum attainable interaction strength $\phi_{\text{max}}$ for the experiment.\vspace{-0.2cm}}
\label{fig:fig2}
\end{figure}
\section{Non-destructive Electron Beam Number Measurements}
We apply the predicted entanglement between spin precession and electron beam number to propose several non-destructive protocols on the electron beam with post-interaction readout purely on the qubit. Measurements along a single axis of the spin qubit allow one to discriminate different electron beam number distributions and measurements along two different spin qubit axes enable recovery of the electron beam's distribution $p(n)$ by sampling different interaction strengths $\phi$. Lastly, through iterative projective measurements on the qubit, one may directly shape the electron beam number distribution. In contrast to quantum sensing protocols that measure particle contributions to the field, these protocols directly interface with the electron beam through the quantum number operator~\cite{zhou2025entanglement, jones2009magnetic}.

\textit{Electron Distribution Discrimination.} For both electron distribution discrimination and determination, we consider the experiment in Figure~\ref{fig:fig2}a. A spin is prepared in the $\ket{\downarrow}$ state and the free electron flies by, applying the scattering matrix from Eq.~\ref{eq: NonRes}. Since the spin rotation is entangled to the number operator on the electron beam, the readout of $\hat{\sigma}_z$ on the spin qubit depends on number of electrons in the bunched beam. Rotating about the $x$-axis prior to readout measures $\hat{\sigma}_y$. Specifically,
$$\langle\hat{\sigma}_z\rangle = \sum_n p(n) \cos(2\phi n), \hspace{0.25in} \langle\hat{\sigma}_y\rangle = \sum_n p(n) \sin(2\phi n)$$
Here, $\phi$ determines how easily this measurement distinguishes different beam distributions. A natural analogy to coherent spin driving arises: $n\phi$ gives an effective Rabi drive rate, while the beam distribution spread induces spin qubit decoherence.

For small $\phi$, such as that attained in the free space interaction with impact parameter $r_\perp = 1~\text{nm}$, the spin operator readout remains near its initial value, producing no discernible differences for different beam distributions, even with mean electron number (current) on the order of hundreds of electrons (Figure 2b, left). However, with a sufficiently strong $\phi$, enhanced by a cavity, discernible differences in qubit readout arise even for mean electron number on the order of tens of electrons. As shown in Figure (2b, right), a pure Fock state electron distribution leads to perfect Rabi flops with increasing mean electron number, while a classical Poissonian distribution yields damped Rabi oscillations. Interaction with an electron beam seeded by bright squeezed vacuum (BSV) driven photoemission -- useful for probing highly nonlinear ionization processes~\cite{rasputnyi2024high, heimerl2024multiphoton, heimerl2025quantum} -- yields an exponential decay in readout signal, due to the BSV distribution's broad support for high electron numbers. Figure~\ref{fig:fig2}c quantifies the relationship between the Fano number, which characterizes how closely a distribution resembles a Poissonian, and the readout of $\langle\hat{\sigma}_z\rangle$.

\textit{Electron Distribution Determination.} Using measurements on both $\langle\hat{\sigma}_z\rangle$ and $\langle\hat{\sigma}_y\rangle$, we may compute
\begin{align}
\hat{N}(\phi) = \langle\hat{\sigma}_z\rangle - i\langle\hat{\sigma}_y\rangle = \sum_n p(n) e^{-i2\phi n},
\end{align}
which is the discrete Fourier transform~\cite{winograd1978computing} of the electron number distribution $p(n)$. Varying the values of $\phi$, done by adjusting beam impact parameter $r_\perp$, recovers $p(n)$ via an inverse Fourier transform on samples of $\hat{N}(\phi)$. Analogous to discrete signal processing, where higher sampling rates are used to resolve fast time-varying signals, a higher interaction strength $\phi$ is required to resolve sharp changes in the $p(n)$. With an effective coupling strength of $\phi = \pi$, the probability distribution is exactly recoverable. However, even with $\phi_\text{max} = 0.1$, we are able to recover distributions with some fidelity, measured by the Kullback-Leibler (KL) divergence, shown in Figure 2d.  By guiding the electron beam back into the qubit-cavity system without measuring the qubit, one may enhance the effective interaction strength, with seven round trips demonstrated experimentally~\cite{seidling2024resonating}, and more trips feasible with improved electron beam guiding.

\begin{figure}
\includegraphics[width=0.45\textwidth]{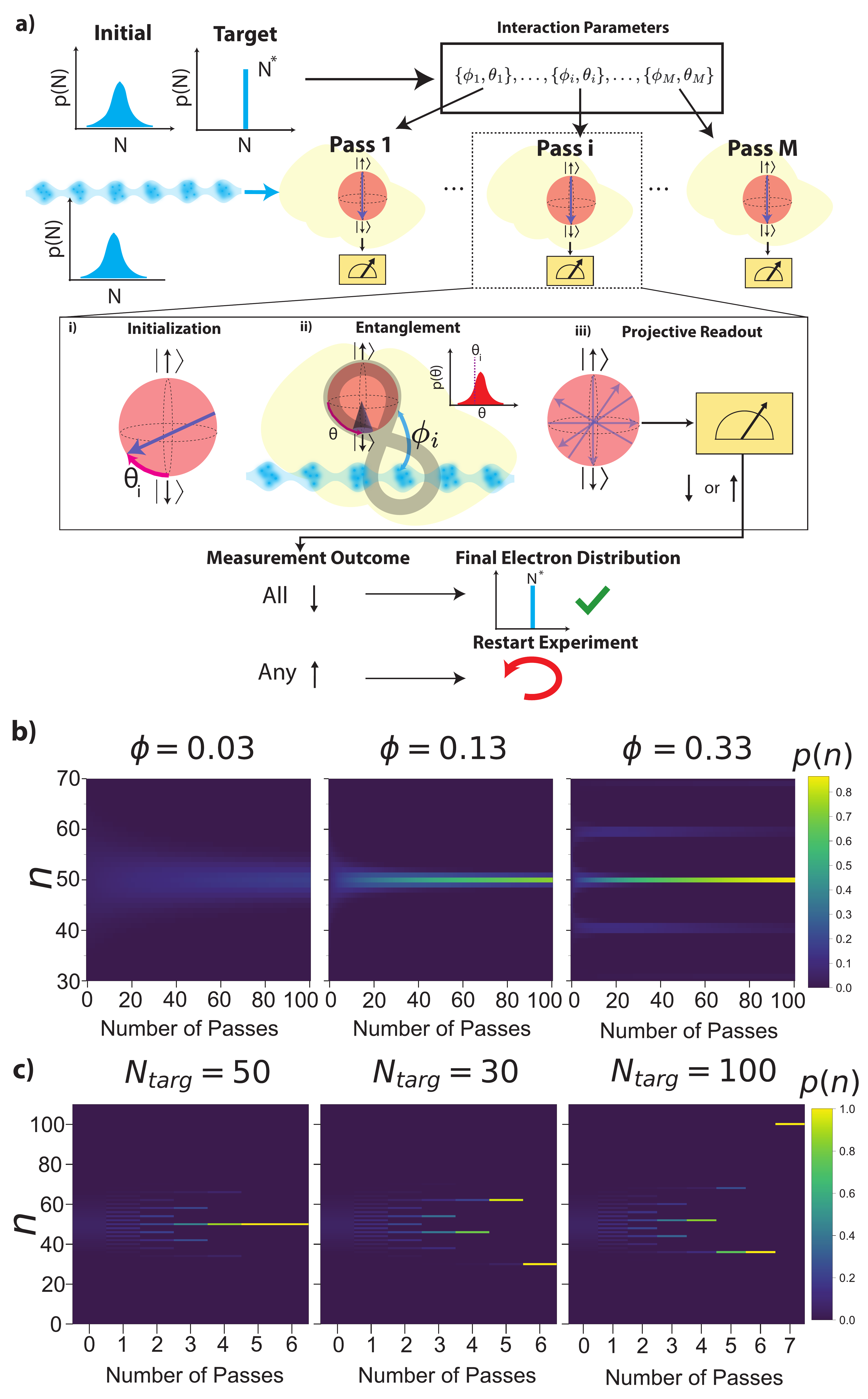}
\caption{\textbf{Electron Number Projection }\textbf{a.}  Protocol for non-destructive electron number projection. The initial and target distribution set the experimental parameters $\phi_i, \theta_i$ for a series of electron beam-qubit interactions. At each interaction, the following are performed: \textbf{i)} Qubit initialized in $\ket{\downarrow}$ and rotated by angle $-\theta_i$. \textbf{ii)} Electron-beam interaction induces number-dependent spin precession. The targeted electron number rotates the spin back to $\ket{\downarrow}$. \textbf{iii)} Projectively measure the qubit in the $\ket{\downarrow}, \ket{\uparrow}$ basis. If all spins are measured in $\ket{\downarrow}$, the electron beam is successfully projected in the target state. \textbf{b. } Projection protocol with uniform interaction parameters $\phi_i$ and $\theta_i = -\phi_i n^*$. The electron beam is initially Poissonian with mean $\mu = 50$ and target state is $\ket{n^*} = \ket{50}$. \textbf{c. }Optimal protocol to reach $\ket{n^*}$ with minimal number of scattering interactions. The electron beam is initially Poissonian with mean $\mu = 50$. The interaction and preparation parameters vary as $\phi_i = \pi/2^i$ and $\theta = n^*\phi_i$.}
\label{fig:fig3}
\end{figure}

\textit{Electron Number Projection.} Using the entanglement between the spin precession and beam number, a series of projective qubit measurements can shape the electron beam to a specific Fock state $\ket{n^*}$, shown schematically in Figure 3a. We send an electron beam, initially in state $\rho_\text{el} = \sum_{n,n'}\rho_{n,n'}\ket{n}\bra{n'}$, through $M$ spin qubit interactions with strengths $\phi_i$. Prior to launching the beam, we apply a rotation of angle $-\theta_i$ along the spin's axis of beam-induced precession. After interaction with the beam, the spin is measured in the $\ket{\downarrow}, \ket{\uparrow}$ basis. If the readout yields $\ket{\downarrow}$, the beam's density matrix becomes 
\begin{align}
\begin{split}
    \rho_\text{el} \to \rho'_\text{el} = \sum_{n,n'} \frac{\cos(\phi n-\theta)\cos(\phi n'-\theta)\rho_{n,n'}}{\sum_{m}\cos^2(\phi m-\theta)p(m)}\ket{n}\bra{n'},
\end{split}
\label{eq:update_rule_rho_unif_int}
\end{align}
where $p(n)=\rho_{n,n}$. Setting $\phi_i n^* = \theta_i$ maximizes the weight of $p(n^*)$ in the modified distribution and minimizes the contributions of $p(n\neq n^*)$. For an experiment where $\theta$ and $\phi$ are fixed for all $i$, the final distribution converges to the targeted one, with the rate of convergence determined by $p(n^*)$ and $\phi$, shown in Figure 3b for a few different choices of $\phi$. However, as shown in Figure 3c, we may engineer different final distributions on the beam number and accelerate convergence by careful selection of different $\phi_i$ parameters. By setting $\phi_i = \frac{\pi}{2^{i}}$ and $\theta_i = -\phi_i n^*$ for $i = 0,...,M-1$, the distribution $p'(n)$ rapidly converges to $\delta_{n,n^*}$, within $\log(\bar{N})$ interactions, where $\bar{N}$ is the support of the initial distribution. This scheme leads to an advantageous scaling in experimental resources required to project macroscopic electron beams.

\section{Conclusion}
Through the formalism described here, we provide a unified framework to study the interactions between electron beams and bound spin and electric dipole qubits in a variety of electromagnetic environments. This allows us to improve the naturally weak interaction between electron beams and bound-electron systems by modifying the underlying environment, providing a solution to one of the primary experimental roadblocks in quantum coherent electron beam physics. We use our framework and the improved interaction strength to propose experimentally viable procedures that can non-destructively measure and manipulate the electron beam's number distribution. Just as photonic state shaping plays a crucial role in cavity and circuit quantum electrodynamics, we anticipate that quantum-controlled electron beam states may serve analogous applications for quantum coherent electron beams and quantum technologies that leverage the spatial resolution of electron beams. While the formalism developed here considers the Markvoian dynamic regimes, where coupling strengths are weak, it is still applicable for both quantum sensing and quantum control regimes, and we anticipate that this theory may be extended to the strong coupling regime and enable the exploration of even richer physics.\\

\begin{acknowledgments}
We acknowledge Joonhee Choi, Zhexin Zhao, Ido Kaminer and Dominic Catanzaro for illuminating discussions related to the presented work.

J.M.G acknowledges support from a Hertz Foundation Graduate Fellowship. A.K is supported by the VATAT-Quantum Fellowship by Israel Council for Higher Education, the Urbanek-Chodorow postdocaral fellowship by the Department of Applied Physics at Stanford University, the Zuckerman STEM leadership postdoctoral program and the Viterbi fellowship by the Technion.  C.R.C is supported by a Stanford Science Fellowship. T.K.L is supported by a NSF Graduate Research Fellowship. S. F. acknowledges a Simons Investigator in Physics grant from the Simons Foundation (Grant No. 827065). TKL and JV are supported by the Vannevar Bush Faculty Fellowship from the US DoD.

\end{acknowledgments}

\bibliography{magnets}
\end{document}